\documentclass[aps,prd,twocolumn,groupedaddress,bibnotes,showpacs]{revtex4}
\usepackage{epsfig}

\newcommand{\povo}{Dipartimento di Fisica, Universit\`{a} di Trento, 
and I.N.F.N., Gruppo di Trento, 38123 Povo (TN), Italy}
\newcommand{\mitma}{Massachusetts Institute of Technology, LIGO, Cambridge, MA 02139}
\newcommand{\uw}{Center for Experimental Nuclear and Particle Astrophysics, University 
of Washington, Seattle, WA 98195}
\newcommand{\uf}{Department of Physics, University of Florida, Gainesville, FL 32611}

\newcommand{\rthz}{\ensuremath{/\mathrm{Hz}^{1/2}}}

\newcommand{\figr}[1]{Fig.\ \ref{#1}}
\newcommand{\eqr}[1]{Eqn.\ \ref{#1}}

\begin{document}


\title{Brownian force noise from molecular collisions and the sensitivity 
of advanced gravitational wave observatories}
\author{R.~Dolesi}
\author{M.~Hueller}
\author{D.~Nicolodi}
\author{D.~Tombolato}
\thanks{Current address: Museo delle Scienze, 38122 Trento, Italy}
\author{S.~Vitale}
\author{P.~J.~Wass}
\thanks{Current address: Imperial College, London, UK}
\author{W.~J.~Weber}
\affiliation{\povo}

\author{M.~Evans}
\author{P.~Fritschel}
\author{R.~Weiss}
\affiliation{\mitma}

\author{J. H. Gundlach}
\author{C. A. Hagedorn}
\author{S. Schlamminger}
\thanks{Current address: 
National Institute of Standards and Technology, Gaithersburg, Maryland 20899, USA}
\affiliation{\uw}

\author{G. Ciani}
\affiliation{\uf}

\author{A. Cavalleri}
\affiliation{Istituto di Fotonica e Nanotecnologie, C.N.R.- Fondazione Bruno
Kessler, 38123 Povo (TN), Italy}

\date{\today}

\begin{abstract}
We present an analysis of Brownian force noise from residual gas damping
of reference test masses as a fundamental sensitivity limit in small
force experiments. The resulting acceleration noise increases
significantly when the distance of the test mass to the surrounding
experimental apparatus is smaller than the dimension of the test mass
itself. For the Advanced LIGO interferometric gravitational wave
observatory, where the relevant test mass is a suspended 340~mm diameter
cylindrical end mirror, the force noise power is increased by roughly a
factor 40 by the presence of a similarly shaped reaction mass at a
nominal separation of 5~mm. The force noise, of order 20~fN\rthz\ for $2
\times 10^{-6}$ Pa of residual H$_2$ gas, rivals quantum optical
fluctuations as the dominant noise source between 10 and 30~Hz. We
present here a numerical and analytical analysis for the gas damping
force noise for Advanced LIGO, backed up by experimental evidence 
from several recent
measurements. Finally, we discuss the impact of residual gas damping on the
gravitational wave sensitivity and possible mitigation strategies. 
\end{abstract}


\maketitle

\section{Introduction}

The Brownian motion of a macroscopic test mass is a fundamental limit in
many small force measurements, which ideally require a test mass (TM)
to be free of stray forces, and thus a reference of purely inertial motion,
except for any known and calibrated suspension forces. The power
spectrum of the Brownian force noise acting on the TM is related to any
source of mechanical dissipation, given by the mechanical impedance
$Z \left( \omega \right)$, through the fluctuation-dissipation theorem
\begin{equation}
S_F \left( \omega \right) 
= 
4 k_B T \, \mathrm{Re} \left[ Z \left( \omega \right) \right] 
= 
4 k_B T \, \mathrm{Re} 
\left( -\frac{F \left( \omega \right)}{ v \left( \omega \right)} \right)
\:\: .
\label{fluc_diss_eqn}
\end{equation}


Gas damping in the molecular flow regime is known to produce viscosity
proportional to the residual gas pressure, $Z \left( \omega \right) =
\beta \sim p$\cite{christian_vacuum,pantolovsky,saulson_thermal_prd,utn_inf_volume},
with $\beta$ referred to here as the gas damping coefficient. In the
limit to which the collisions can be treated as independent impulses,
the resulting Brownian force noise has a frequency independent spectrum.
Residual gas force noise has recently been reconsidered in the context
of gravitational experiments with geodesic-reference TM \cite{utn_prl_damp,
UW_squeeze_damping}. In these and other experiments, the proximity of
surrounding surfaces can significantly increase the gas damping, and
thus the resulting force noise, over that observed for the same TM in an
infinite volume filled only with gas. This phenomenon, referred 
to here as proximity-enhanced gas damping, has also been observed
and studied for MEMS oscillators, under the name of squeeze film damping
\cite{bao_squeeze,suijlen_squeeze}.

While there is a continuum of behavior between the free damping (infinite 
gas volume) limit and proximity damping, we will divide the total
molecular impact force noise into these contributions,
\begin{equation}
S_F \left( \omega \right) 
= S_F^{\infty} + \Delta S_F \left( \omega \right)  \:\: .
\label{eqn_def}
\end{equation}
$S_F^{\infty} = 4 k_B T \beta^{\infty}$ is the noise for the TM in
an infinite gas volume, while the excess
$\Delta S_F$ depends on the TM proximity to 
surrounding surfaces.  

Though long unnoticed in experimental gravitation, proximity-enhanced
gas damping has a straightforward physical explanation: motion of a TM,
with characteristic dimension $s$, in the vicinity of another surface,
at some small distance $d$, creates a transient squeezing of any
residual gas in the gap between the two bodies. With the subsequent
molecular flow out of the gap, there is an associated pressure drop,
proportional to the molecular current, and thus a velocity dependent
force. In an alternate picture, the Brownian force noise increases
because of the correlation between repeated impacts of a molecule on the
TM as it stochastically moves along the gap. As the same random walk
statistics govern the molecular flow down the channel, these two
pictures -- of dissipative flow and correlated collision impulses -- are
equivalent, as \eqr{fluc_diss_eqn} requires. 

The grouping of repeated impulses into ``macro-collisions'' establishes
a natural interaction (or correlation) time $\tau$, the typical time
needed for random walk molecular diffusion out of the interbody gap.
This creates a high frequency cutoff, $\omega \approx \tau^{-1}$, above
which the force noise decreases. In terms of the macroscopic flow
impedance picture, TM motion for $\omega \tau \gg 1$ is too fast to
allow molecular flow, with the real, dissipative, part of $Z \left(
\omega \right)$ decreasing to leave only the imaginary impedance of gas
compression. We will return to these simple arguments later in the
article for an approximate analytical model that is useful in
interpreting the results of our study, for $\Delta S_F$ and $\tau$ in
terms of the aspect ratio $s/d$. 

Recent experimental studies \cite{utn_prl_damp,UW_squeeze_damping} place
the physics of proximity-enhanced gas damping on a quantitative footing.
These studies both measure the gas damping on the free oscillation of
torsion pendulums, focusing on the low frequency range $\omega \tau \ll
1$ where $\beta$ is independent of frequency. The torsion pendulum
inertial elements in these studies were a cubic TM inside a cubic
enclosure and a rectangular plate in close proximity of a second
parallel plate. The TM size-to-gap aspect ratios studied range from
roughly 4 to 400. Both experiments observed a clear proximity excess 
in the gas
damping, well beyond calculated values of $\beta^{\infty}$. These
experiments will be discussed in Sec. \ref{sec_exper}. 

\begin{figure}[tb]
\includegraphics[scale=1]{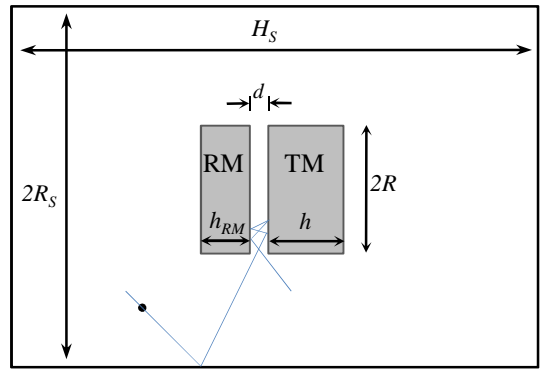}
\caption
{Cartoon illustration of the Advanced LIGO experimental configuration
studied and simulated here, seen perpendicular to the symmetry axes of
the cylindrical test (TM) and reaction (RM) masses, with radii $R$ =
17~cm and heights $h$ = 20~cm and $h_{RM}$ = 13~cm. The gap $d$,
nominally 5~mm for Advanced LIGO, is varied in the simulations. The
simulated experimental enclosure, not shown to scale here, is a coaxial
cylinder of radius $R_S = 10 R$ and length $H_S = 11 h + h_{RM} + d$.}
\label{fig_sim}
\end{figure}

Gas damping has been analyzed as a force noise source for terrestrial
gravitational wave interferometers \cite{saulson_thermal_prd} but should
be reconsidered for the next generation experiments, both for their
stringent acceleration noise requirements and for the close proximity of
TM to the surrounding apparatus. For Advanced LIGO
\cite{harry_adv_ligo,adv_ligo_ref_design}, 
the envisioned end mirror TM are right cylinders
of fused silica with radius $R$ = 17~cm and length $h$ = 20~cm (mass $M
\approx$ 40~kg). In an infinite gas volume, at $T$ = 293~K and 
pressure $p = 2 \times 10^{-6}$~Pa of residual H$_2$ gas
($m_0$ = 2~amu), the molecular impact force noise, integrated over 
the entire cylindrical TM surface, would be
\cite{utn_inf_volume}
\begin{eqnarray}
S_F^{\infty} 
& = & 
p \left(128 \pi m_0 k_B T \right)^{1/2} R^2 
\left( 1 + \frac{h}{2R} + \frac{\pi}{4} \right)
\nonumber \\
& \approx &
\left[ 3.2 \: \mbox{fN/Hz}^{1/2} 
\times \left( \frac{p}{2 \times 10^{-6} \: \mbox{Pa}} \right) ^{1/2}
\right]^2
\:\: .
\label{S_F_inf} 
\end{eqnarray}
This free damping compares favorably with the Advanced LIGO target,
which requires total force noise below 40~fN\rthz\ at frequencies near
30 Hz to reach a strain sensitivity of $10^{-23}$\rthz\ (equivalent to a
TM acceleration noise of roughly 1 fm/s$^2$\rthz). However, the original
Advanced LIGO design includes cylindrical ``reaction
masses,'' with the same radius and 13~cm length, facing the TM at a
nominal separation $d$ = 5~mm (see \figr{fig_sim}). The close reaction
masses are needed for all four cavity mirror TM, for electrostatic force
actuation and for compensating optical thermal lensing from TM laser heating. 
The relevant aspect ratio of $2R/d \approx 70$ is
well into the regime of proximity-enhanced gas damping, and thus
the problem merits a thorough understanding.


In this article, we calculate the force noise and diffusion time using a
numerical simulation of a molecule exchanging momentum with the TM
mirrors in the Advanced LIGO geometry, as a function of the gap size
(Section \ref{sec_numerical}). In Section \ref{sec_analytical} these
results are compared with an approximate analytical model, which
successfully explains the functional dependence of the excess force
noise on gap size in terms of a simple random walk picture. In Section
\ref{sec_exper}, the physics and simulation techniques of Secs.
\ref{sec_numerical} and \ref{sec_analytical} are compared with
experimental evidence from similar systems. 
Finally, in Section \ref{sec_discussion_results}
we address the impact of gas damping force noise on the sensitivity of
advanced gravitational wave detectors and discuss possible mitigation
strategies.  While the article focuses on the Advanced LIGO geometry,
we will discuss design consideration for other advanced gravitational
wave observatories in the conclusion.  
  
As in previously cited experimental and numerical studies
\cite{utn_prl_damp,utn_inf_volume,UW_squeeze_damping,suijlen_squeeze},
we consider completely inelastic molecular collisions with the TM, with
subsequent immediate random reemission with a cosine angular
distribution. The inelastic, diffuse scattering hypothesis is verified
experimentally for spinning rotor pressure gauges and is the basis for
their use as absolute pressure calibration standards
\cite{fremerey_j_vac_sci,dittmann_j_vac_sci}. This also guarantees time
reversal invariance and Maxwell-Boltzmann statistics in the gas phase
(see Ref. \cite{comsa}). The immediate reemission approximation is valid
in the limit that the effective sticking time, $\tau_{st}$, is
negligible compared to that needed to cross the gap between TM,
$\frac{d}{v_T}$. This is justified for H$_2$ and other possible gas
species on the TM SiO$_2$ surface at room temperature, based on the
relevant adsorption potentials, in Appendix \ref{app_sticking_time}.
Finally, we treat the average gas pressure as uniform in the system,
which requires that the gap pressure is not dominated by local
outgassing from the test or reaction mass, which would cause a stable
increase in the relevant local pressure. We return to these last two
points in the conclusions. 

\section{Numerical study of Brownian force noise}
\label{sec_numerical}

We have performed Monte Carlo simulations to analyze the gas damping
noise for the Advanced LIGO geometry. The simulations (following those
performed in \cite{utn_prl_damp}) trace the trajectory of a gas molecule
as it moves inside a large volume containing the cylindrical test mass
and reaction mass. The molecular species is chosen to be H$_2$ at room
temperature, the dominant residual gas expected in the Advanced LIGO vacuum
chamber, which sets the scale of momentum exchange upon impact and the
time scale between collisions through the characteristic thermal
velocity $v_T \equiv \sqrt{k_B T / m} \approx 1100$ m/s. The volume
enclosing the two cylinders, ideally infinite, is chosen to be a
cylinder ten times larger, in both radius and length, than the volume
envelope defined by the two cylinders and the intervening gap. 

The simulation starts by random selection of a molecular position, from
a uniform distribution in the available volume -- including the gap but
also the much larger, and thus more probable, space surrounding the
cylinders -- and velocity, from a Maxwell-Boltzmann distribution. The
impacts with the walls are completely inelastic, with subsequent
immediate random thermal reemission following a cosine angular
distribution, independent of the incoming velocity, with the resulting
gas distribution obeying Maxwell-Boltzmann statistics (the reemission
distribution is given in \eqr{app_distrib}, as in Ref.
\cite{utn_inf_volume}). At each collision with the TM --regardless of
whether the molecule strikes on the TM surface facing the gap or on the
opposing side or outer wall -- the time and exchanged momentum vector
are recorded. Simulations are run, for a single molecule, for a time
$T_0$ (in the range of 10$^{-6}$ - 10~s, chosen to abundantly cover the
full range of the observed values of $\tau$) and then repeated for many
trials. Finally, the simulations have been performed as a function of
gap $d$ between the two cylinders, in the range from 100~$\mu$m to 1~m. 

One way to probe the simulation is to observe the statistical
fluctuations of the total momentum $\Delta q$ exchanged along the
critical cylindrical symmetry axis in repetitions of the simulations of
duration $T_0$. This allows simple extraction of the parameters
describing the amplitude and spectral shape of the force noise. For a
process with white force noise up to some high frequency rolloff
associated with a time constant $\tau$, with $S_F = \frac{S_{F_0}}{1 +
\left( \omega \tau \right)^2}$, one can calculate (see Appendix
\ref{app_calc_dp2})
\begin{equation}
\langle \left( \Delta q \right)^2 \rangle = 
\frac{S_{F_0} T_0}{2} 
\left( 1 - \frac{1 - \exp \left( - \frac{T_0}{\tau}\right)}
{\frac{T_0}{\tau}}  \right)  
\: \: ,
\label{eqn_exp_delta_p_2}
\end{equation} 
with $\langle \left( \Delta q \right)^2 \rangle = \frac{S_{F_0}T_0}{2}$
in the limit $T_0 \gg \tau$.  

In our simulations, we directly calculate the average scatter in $\Delta
q$ from the total momentum exchange in the different trials, typically
calculating $\langle \left( \Delta q \right)^2 \rangle$ from groups of
10$^3$ (for the longer simulations) to 10$^6$ (shorter $T_0$) single
molecule experiments and then estimate the simulation uncertainty
between 100 such groups. In a physical experiment with many
non-interacting gas molecules, $\langle \left( \Delta q \right)^2
\rangle$, and thus also the force noise power, is proportional to the
number of molecules in the system, allowing us to scale our single
molecule simulations to a given pressure (we choose $2 \times 10^{-6}$~Pa 
as a reference pressure to put a relevant experimental 
scale to the data presented). 

\begin{figure}[tb]
\includegraphics[scale=0.66]{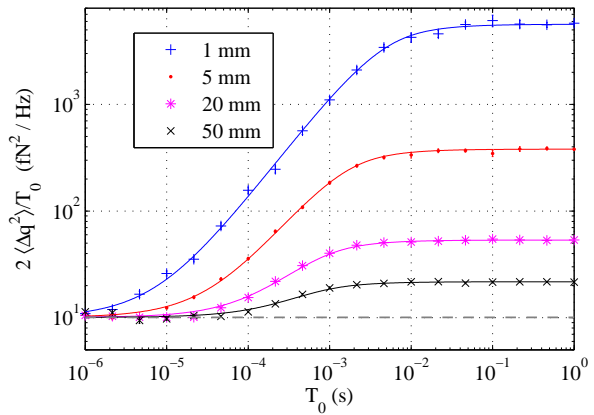}
\caption
{Simulation data for the scatter in the total momentum $\Delta q$
transferred to the TM in simulation time $T_0$, for several values of
the intercylinder gap $d$, each fit to the model in
\eqr{eqn_dp2_fit_model}. The range of gaps studied represents 
aspect ratios (2$R$/$D$) from nearly 7 to 340.  
The dashed gray line represents the prediction
in the limit of infinite gap, $S_{F}^{\infty}$, which is recovered for
very short simulation times. Data are
scaled for a reference pressure $p$ = 2 $ \times 10^{-6}$ Pa.}
\label{dp2_T0}
\end{figure}

Figure \ref{dp2_T0} shows simulation data for $\frac{2 \langle \left(
\Delta q \right)^2 \rangle }{T_0}$ as a function of measurement time
$T_0$ for several values of the gap $d$. The data show a clear increase
in force fluctuations with decreasing gap. The time dependence shows
saturation for large $T_0$, as predicted by \eqr{eqn_exp_delta_p_2} and
which we can associate with $\tau$, but also a saturation to a nonzero
value at short $T_0$, for which this simple model with a single high
frequency rolloff does not account. This low-$T_0$ saturation has,
however, a simple physical explanation: for $T_0$ below roughly $d /
v_T$ -- the time necessary for a molecule to cross the intercylinder
gap, see Appendix \ref{app_calc_r2_t} -- a molecule will hit the TM at
most once during the simulation. In this limit the simulation becomes
blind to any correlation effects related to proximity, as for a
cylindrical TM in an infinite volume.  In fact, the low-$T_0$ saturation
values shown in \figr{dp2_T0} converge, for different values of $d$, to
the infinite volume value, $S_F^{\infty}$.  There would also be a natural
time scale and associated high frequency rolloff relevant to the
simulation in the single collision (or infinite volume) limit, but it
corresponds to the single collision duration, which the simulation takes
to be infinitesimal. These curves thus both confirm the infinite volume
force noise prediction for short simulations and, in the longer
simulations, indicate an increasing excess force noise for decreasing
$d$. 

We note that no excess force fluctuations are observed for the
components orthogonal to the cylinder symmetry axis, for which the
infinite volume limit of Ref. \cite{utn_inf_volume} is verified. 
The correlation in the force from successive impacts is
only non-zero for the component normal to the surface, for which the
direction of the impulse is always toward the test mass center, whereas
the sign of the shear component of momentum exchange along an axis
parallel to the surface is random. From a macroscopic damping
standpoint, TM motion parallel to the reaction mass does not
squeeze the gas between the two bodies, and thus creates no dissipative
gas flow in the narrow gap.

\begin{figure}[tb]
\includegraphics[scale=0.66]{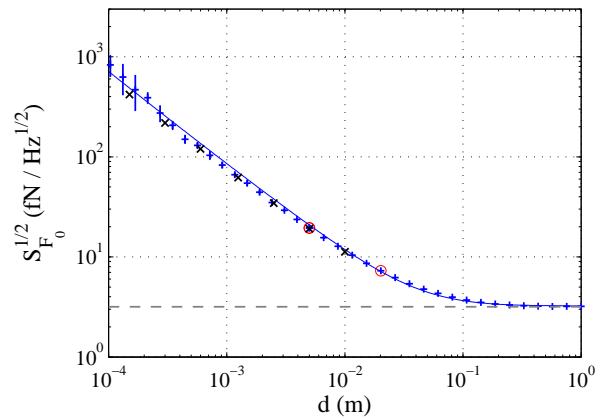}
\caption
{Simulation data (blue dots) for the force noise linear spectral density
as a function of the gap, $d$, for $p$ = 2 $\times 10^{-6}$~Pa, with a
fit to the simple analytical model (\eqr{model_fit}). The dashed gray
line is the theoretical infinite gap limit. Points at $d$ of 5 and 20 mm
are circled to indicate, respectively, the geometries of the baseline
and alternative designs discussed in the conclusion. Points shown as black
``$\times$'' are simulation data for the molecular escape time,
converted into force noise using the model in Sec.
\ref{sub_sec_piston}.}
\label{S_F_d}
\end{figure}

\begin{figure}[t]
\includegraphics[scale=0.66]{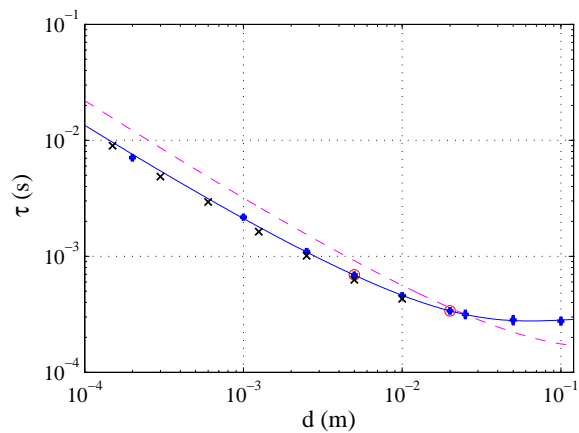}
\caption
{Value of the characteristic time constant $\tau$ extracted from the
simulation force fluctuations (blue dots), as a function of gap $d$,
using the model in \eqr{eqn_dp2_fit_model}. The dashed pink curve
represents the raw result of the approximate analytic model
(\eqr{calc_tau}) in Sec.\ref{sub_sec_tau}, while the solid blue curve
represents a two parameter fit to the same model, discussed in the text.
Again, circled points at 5 and 20~mm indicate the baseline and alternate
geometries. Points marked with a black ``$\times$'' are simulation
values for the average molecular escape time from the intercylinder gap,
which correspond approximately to the values of $\tau$ extracted from
the force fluctuations. }
\label{tau_d}
\end{figure}

With these results in mind, we modify the force noise model to 
\begin{equation}
S_F \left( \omega \right) 
= 
S_F^{\infty} 
+ 
\frac{\Delta S_{F_0}}{1 +\left( \omega \tau \right)^2}
\: \: ,
\label{force_noise_model}
\end{equation}
where $\Delta S_{F_0}$ represents the excess low frequency force noise
associated with the proximity effect. For each $d$, the curves in
\figr{dp2_T0} are thus fitted to a model modified from
\eqr{eqn_exp_delta_p_2} to include the infinite volume limit,
\begin{equation}
\frac{2 \langle \left( \Delta q \right)^2 \rangle}{T_0} = 
S_F^{\infty} + 
\Delta S_{F_0}  \left( 1 - \frac{1 - \exp \left( - \frac{T_0}{\tau} \right)}
{\frac{T_0}{\tau}}\right)  
\: \: ,
\label{eqn_dp2_fit_model}
\end{equation}
with the total low frequency force noise $S_{F_0}$ obtained in the limit
$T_0 \gg \tau$, 
\begin{equation}
\lim_{T_0 \rightarrow \infty} 
\frac{2 \langle \left( \Delta q \right)^2 \rangle}{T_0}
\equiv S_{F_0} 
= 
S_F^{\infty} + \Delta S_{F_0} 
\: \: .
\label{eqn_dp2_limit}
\end{equation}
This allows extraction of the free fit parameters corresponding to the
low frequency force noise $S_{F_0}$ and diffusion time $\tau$, which are
plotted as a function of $d$ in, respectively, Figs.\ \ref{S_F_d} and
\ref{tau_d} \footnote{In practice, the data for $S_{F_0}$ in
\figr{S_F_d} have been extracted directly from the scatter in the
momentum in dedicated long simulations at fixed $T_0$ = 10~s for many
values of $d$, using the limit of \eqr{eqn_exp_delta_p_2} for $T_0 \gg
\tau$, $S_{F_0} = \frac{2 \langle \left( \Delta q \right)^2
\rangle}{T_0}$.}.

\begin{figure}[tb]
\includegraphics[scale=0.66]{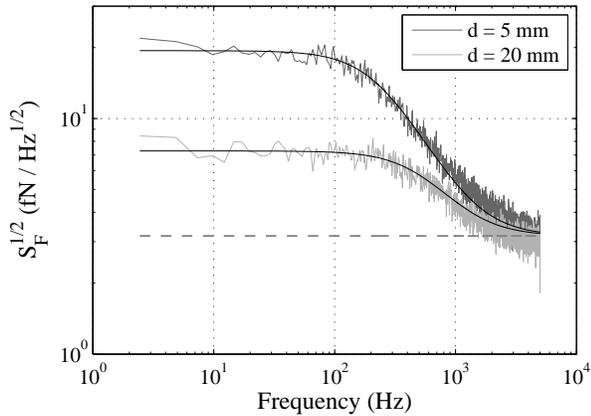}
\caption
{Simulation force noise linear spectral density as a function of
frequency for 5 and 20~mm gaps, using $p = 2 \times 10^{-6}$~Pa. Also
plotted, with solid black lines, is the model from \eqr{force_noise_model},
using the parameters extracted from the data in \figr{dp2_T0}. The
dashed gray line corresponds to the force noise in the infinite volume
limit.}
\label{spectrums}
\end{figure}

The Monte Carlo molecular dynamics simulation can also be probed by
sampling the momentum exchange with the test mass in repeated time
intervals to create a force time series, which can then be analyzed for
the frequency dependent noise spectrum. Force noise spectra have been
calculated using a 100 $\mu$s sampling time -- and thus well shorter
than the relevant values of $\tau$ -- for a simulation of 10 seconds,
which has been analyzed in overlapping 0.4~s windows, with averaging
over roughly 25 statistically independent spectra. These data, displayed
for 5 and 20~mm gaps in \figr{spectrums}, are in agreement with the model
in \eqr{force_noise_model}, which has been plotted on top of the noise
spectrum data using the values of $\Delta S_{F_0}$ and $\tau$ obtained
from $\langle \left( \Delta q \right)^2 \rangle $ as described above and
presented in Figs. \ref{dp2_T0} - \ref{tau_d}. 

A variation on this simulation can be used to extract the average escape
time for a molecule inside the intercylinder gap. Here, the initial
position of the molecule is distributed uniformly inside the cylindrical
space between TM and RM, with the simulation run until the molecule
exits this gap. This has been performed for 7 values of $d$, each time
averaging over 10000 trial molecules. These data are plotted with the
values of $\tau$ extracted from the force fluctuations in \figr{tau_d}.
The agreement, at the 10-15\% level, with the time constant relevant to
the force noise spectrum rolloff, confirms the intuitive interpretation
of diffusion time discussed in the introduction. Additionally, as will
be discussed in the following section, the diffusion time $\tau \left( d
\right)$ is linked directly, with a simplified analytical model, to an
estimate of the resulting force noise. This projection for the white
force noise level is shown here in \figr{S_F_d}.

Before discussing the quantitative impact of the observed noise increase
on the gravitational wave detection sensitivity, we turn first to a
discussion of the physics of excess gas damping in the context of
simplified analytical models and then to the experimental evidence
underpinning the simulation and predictions. 

\section{Comparison with simplified analytical models}
\label{sec_analytical}
A combination of simple analytical models is useful for illuminating 
the physics of proximity-enhanced gas damping, but also for arriving at a
closed form approximate expression for the excess force noise as a 
function of the gap $d$.  

\subsection{Macroscopic damping for a piston weakly connected to an
external gas reservoir}
\label{sub_sec_piston}
A simple macroscopic model for proximity-enhanced gas damping, following
the discussion in the introduction and Refs.
\cite{UW_squeeze_damping,suijlen_squeeze}, comes from approximating the
intercylinder gap as an isothermal volume at uniform pressure connected
to the external gas reservoir at pressure $p$ via a molecular flow
impedance, $Z_M$, with units of $\left [ \mathrm{s} / \mathrm{m}^3
\right ]$. In the presence of a displacement $x \left( t \right)$ of the
TM along its symmetry axis, the volume is $V = \pi R^2 \left ( d + x
\right)$, and we can express the pressure as $p + \delta p \left( t
\right)$. We can then write the standard ideal gas and molecular flow
equations for the instantaneous number of molecules $N_0$ in the gap, 
\begin{eqnarray}
k_B T N_0  
& = & 
\left(  p + \delta p \right) \pi R^2 \left( d + x \right)
\nonumber \\  
k_B T \frac{dN_0}{dt} & = & - \frac{\delta p}{Z_M} 
\label{piston_eqns}
\end{eqnarray}
Considering small displacements  $x \ll d$, these expressions can be 
combined into a linearized equation for $\delta p$, 
\begin{equation}
\frac{d \delta p}{dt} + \frac{\delta p}{\tau} 
= 
- \frac{p}{d} \frac{dx}{dt}
\:\: ,
\label{eqn_delta_p}
\end{equation}
where we have reexpressed the molecular impedance in terms of 
the effective pressure equilibration time 
$\tau = Z_M V = Z_M \pi R^2 d$, analogous to an $RC$ 
discharge time.

To calculate the mechanical impedance, or relationship of 
force to velocity $v$, \eqr{eqn_delta_p} can be evaluated in the 
frequency domain for harmonic TM motion, using $v = i \omega x$.
The pressure increase $\delta p$ is 
multiplied by surface area to obtain force,
\begin{equation}
F = -\pi R^2 \frac{p \tau}{d} \frac{i \omega x}{1 + i \omega \tau}
=  -v \times \pi R^2 \frac{p \tau}{d} 
\frac{1 - i \omega \tau}{1 + \left( \omega \tau \right)^2}
\:\: .
\label{eqn_impedance}
\end{equation}
Equation \ref{eqn_impedance} describes a mechanical impedance that is
primarily real and thus dissipative at lower frequencies. At higher frequencies,
the impedance becomes imaginary, and thus a lossless spring as gas
compression occurs without sufficient time for gas flow out of the gap.
For the force noise, the fluctuations-dissipation relation in
\eqr{fluc_diss_eqn} can be evaluated with the real part of \eqr{eqn_impedance} 
to yield
\begin{equation}
\Delta S_F 
= 
4 k_B T \times \pi R^2 \frac{p \tau}{d} \frac{1}{1 + \left( \omega \tau \right)^2}
\:\: .  
\label{fluc_diss_eqn_tau}
\end{equation}
This low-pass form of the predicted force noise is indeed confirmed by the 
simulations, in Figs. \ref{dp2_T0} and \ref{spectrums}. 

For the interaction-free regime studied here, the pressure equilibration
time $\tau$ is equivalent to the average time that a molecule needs to
escape from the gap into the surrounding gas reservoir. This escape time
is plotted in \figr{tau_d} and, converted into a low frequency force
noise level using \eqr{fluc_diss_eqn_tau}, \figr{S_F_d}. This 
macroscopic picture is approximate in that it assumes the pressure
inside the gap to be uniform, while physically there must be a radial
pressure gradient in order to drive gas flow out of the gap upon
squeezing. Nonetheless, there is a remarkable agreement for the
cylindrical geometry between the average escape time and the 
simulation force fluctuation values for force noise and time constant $\tau$. 

To complete a predictive analytical model for proximity damping, we
require an independent estimate for the time constant $\tau \left( d
\right)$, which contains the physics of the molecular diffusion process
and is addressed in the following subsection.

\subsection{Random walk calculation of escape time $\tau$}
\label{sub_sec_tau}
The escape time $\tau$ can be estimated by simple random walk arguments
for a molecule in the intercylinder gap. This model is relevant in the
limit $d \ll R$, with a typical molecule making many
collisions before escaping the gap. 

The number of collisions $N$ that a particle makes before diffusing out
of the intercylinder gap can be estimated from the random walk
mean square distance traveled along a direction perpendicular to the
cylinder axis, or $N \langle r^2 \rangle$, where $\langle r^2 \rangle$
is the mean square lateral displacement that a particle makes in a
single flight between opposing faces of the two cylinders. Thus,
covering a distance $R$ to diffuse out of the gap will require $N
\approx \frac{R^2}{\langle r^2 \rangle}$. The characteristic time $\tau$
that a molecule takes to random walk out of the gap will thus be $\tau
\approx N \langle t \rangle$, where $\langle t \rangle$ is the average
time of flight to cross the gap between opposing faces of the two
cylinders.

Estimates of $\langle r^2 \rangle$ and $\langle t \rangle$ can be
calculated analytically for a particle emitted, with cosine-law
distribution, from the center of one of the cylinder faces into the gap
of height $d$ and radius $R$ (see Appendix \ref{app_calc_r2_t}): 
\begin{eqnarray}
\langle r^2 \rangle  
& \approx & 
d^2 \ln \left[ 1 + \left( \frac{R}{d} \right)^2 \right]  
\nonumber \\
\langle t \rangle 
& \approx & 
\left( \frac{\pi}{2} \right)^{1/2} 
\frac{d}{v_T}
\: \: .
\label{RMS_x_t}
\end{eqnarray}
 
We can use these values to estimate $N$, 
\begin{equation}
N 
\approx \frac{R^2}{\langle r^2 \rangle} 
\approx \frac{R^2}{d^2 \ln \left[ 1 + \left( \frac{R}{d} \right)^2 \right]}
\label{calc_N}
\end{equation}
and then the molecular escape time $\tau$, 
\begin{equation}
\tau 
\approx 
N \langle t \rangle
\approx 
\left( \frac{\pi}{2} \right)^{1/2} \frac{ R^2  }
{d v_T \ln \left[ 1 + \left( \frac{R}{d} \right)^2 \right]}
\: \: .
\label{calc_tau}
\end{equation}

Inserting the result of 
\eqr{calc_tau} into \eqr{fluc_diss_eqn_tau} and recalling the 
thermal velocity $v_T \equiv \sqrt{k_B T / m}$, we obtain the low 
frequency force noise
\begin{equation}
\Delta S_{F_0} 
\approx
p \left( 8 \pi m_0 k_B T \right)^{1/2} 
\pi R^2 
\frac{R^2}{d^2 
\ln \left[ 1 + \left( \frac{R}{d} \right)^2 \right] }
\: \: .
\label{force_noise_tau_prediction}
\end{equation}
The force noise thus has a prefactor -- proportional 
to $p$ and surface area $\pi R^2$ -- similar to the 
infinite volume result (\eqr{S_F_inf}), multiplied by a
dimensionless excess factor that depends on the aspect 
ratio via $R/d$.  

This model for $\tau$ and $S_{F_0}$ is also approximate. The true mean
square displacement $\langle r^2 \rangle$ depends on the position on the
cylinder face from which the particle is emitted, as will the expected
full escape time. Additionally, the model treats the random walk as a
Gaussian process, with the mean square step size as the single parameter
characterizing the statistics. For $R/d \rightarrow \infty$ this
approximation is catastrophic, as the variance of the step diverges (see
App. \ref{app_calc_r2_t} and Ref. \cite{UW_squeeze_damping}). However,
for the scope of estimating the diffusion time out of a finite radius
cylindrical volume, the calculation of the step size variance can be
truncated to consider steps no larger than $R$. As seen in earlier
studies\cite{utn_prl_damp} and as will be shown in the next sections,
this is a good approximation to the physics seen in the simulations. 

\subsection{Equivalent shot noise model}
\label{sub_sec_shot}
Independently from the macroscopic piston damping model in Sec.
\ref{sub_sec_piston}, the calculation of $\tau$ in Sec.
\ref{sub_sec_tau} can be used for direct estimate of the force noise
using a shot noise model. Here, the ``shot'' of momentum corresponds not
to the momentum exchange of a single molecular impact, but rather to the
total momentum exchanged by a single molecule over $N$ impacts with the
TM before escaping from the intercylinder gap. 

The force shot noise can be expressed 
\begin{equation}
\Delta S_{F_0} = 2 \left(\delta q \right) ^2 \lambda
\label{shot}
\end{equation}
The characteristic impulse $\delta q$ is the typical thermal momentum
exchange of a single collision ($2 m_0 v_T$) multiplied by the number $N$
of collisions that a molecule makes with the TM on a typical pass through the 
the gap,
\begin{equation}
\delta q \approx 2 N m_0 v_T
\: \: .
\label{dp_impulse} 
\end{equation}
The characteristic rate of such summed impulses, $\lambda$, is the rate
at which molecules enter, and exit, the inter-cylinder gap. This is
the expected number of molecules between the cylinders divided
by the diffusion time $\tau$, 
\begin{equation}
\lambda \approx \frac{p}{k_B T} \times 
\pi R^2 \: d 
\times \frac{1}{\tau} 
\label{lambda}
\end{equation}

Finally, insertion of these values into \eqr{shot} 
gives the force shot noise,  
\begin{eqnarray}
\Delta S_{F_0} 
& \approx &
2 \left(4 N^2 m_0 k_B T \right) 
\times \left( \frac{p}{k_B T} \pi R^2 \, d \frac{1}{\tau}\right)
\nonumber \\
& \approx & 
\frac{4}{\pi}
p \left( 8 \pi m_0 k_B T \right)^{1/2} 
\pi R^2 
\frac{R^2}{d^2 
\ln \left[ 1 + \left( \frac{R}{d} \right)^2 \right] }
\label{shot_noise_prediction}
\end{eqnarray}
This shot noise prediction differs from that of
\eqr{force_noise_tau_prediction} by the numerical factor $\frac{4}{\pi}$
and thus, at the approximate level of this calculation, can be considered
equivalent to \eqr{force_noise_tau_prediction}. 

\subsection{Comparison with simulation results}
Using \eqr{force_noise_tau_prediction} as a model for the excess gas
damping force noise, a fit of the simulation data for $S_{F_0}$ to the
model
\begin{eqnarray}
S_{F_0}  \left( d \right)  
\approx    
A S_F^{\infty} +
\nonumber \\ 
B  p \left( 8 \pi m_0 k_B T \right)^{1/2} 
\pi R^2  \left\{ \frac{R^2}{d^2 \ln \left[ 1  + \left( \frac{R}{d} \right)^2 \right]} 
- 1 
\right\}
\label{model_fit}
\end{eqnarray}
is shown in \figr{S_F_d}. Here, $S_F^{\infty}$ is defined as in
\eqr{S_F_inf}, and the subtraction of unity in the second term is
inserted to give a null contribution for $d \to +\infty$, attributing
all residual damping to the infinite volume term. We obtain $A = 1.038
\pm 0.004$ and $B = 0.78 \pm 0.005$. The value of the excess damping
factor $B$ is of order unity, indicating that that this simplified model
indeed gives the right magnitude of the excess damping, though the observed
agreement at the  25\% level is perhaps better than expected given
the approximations of the model. It is likely that the finite size of
the simulated system is responsible for the several percent excess
observed for the infinite gap limit (coefficient $A$). 

While the fit of this approximate model is not particularly good by
statistical standards, with $\chi^2 \approx 12$, this model nonetheless
gives a useful analytical formula for $S_F$ as a function of gap, with
30\% accuracy in noise power spectral density across four decades of $d$.
We note that fitting to a model that neglects the logarithmic
contribution -- thus considering the excess damping term as $\Delta
S_{F_0} \propto d^{-2}$ -- deviates from the simulation data by more
than a factor 2 across the same range. Similarly unsatisfactory results
are obtained with other power law fits, which can work well over a given
single decade in $d$ but not wider ranges, with the logarithmic slope
observed to change, steepening towards -2 with decreasing gap. This
confirms the basic physics given in Sec. \ref{sub_sec_tau}. 

The values of $\tau$ extracted from \eqr{eqn_dp2_fit_model} are shown in
\figr{tau_d}. These are compared with values associated with the
simulation mean escape time from the gap. Shown as a dashed curve
is a simple, unscaled plot of the approximate analytical estimate of $\tau$
from \eqr{calc_tau} \footnote{Here, however, we insert into
\eqr{calc_tau} the ``exact'' expression for $\langle t \rangle$,
calculated in Appendix \ref{app_calc_r2_t}. This is relevant only at the
largest values of $d$, where the random walk picture starts to break
down, anyways.}, which is within a factor 2 of the simulation results
across the three decades of $d$ for which we extract $\tau$. Also shown,
with a solid curve, is a two parameter fit to the data, using
\eqr{calc_tau} but allowing multiplicative prefactors both to the
overall expression for $\tau$ and as a scale factor for $d$ in the
logarithmic term. This gives a statistically good fit, with the relevant
scale factor effectively lengthening $d$ in the logarithmic term, 
by 3.2 (equivalently, shortening $R$). This is likely related to the
oversimplification of assuming molecules starting from the geometric
center -- and not a range of positions in the intercylinder gap -- in
our calculations of $N$, $\langle t \rangle$, and $\langle r^2 \rangle$.
We note that similar scaling of $d$ in the logarithmic term for the
force noise $S_{F_0}$, discussed in the preceding paragraph, has no
relevant impact on the quality of the fit in \figr{S_F_d}. 

\section{Summary of experimental evidence}
\label{sec_exper}

Two recent experimental efforts provide a verification of the general
physics of proximity-enhanced gas damping and of the simulation
techniques employed in this study. Aside from the specifics of the
geometries studied, the simulations used to analyze these experiments
differ from those described in Sec. \ref{sec_numerical} only in that
they estimate a rotational damping coefficient, $\beta_{rot} =
\frac{S_N}{4 k_B T}$, using the fluctuations in the average torque,
rather than force. Both experimental studies calculate the gas
contribution to $\beta_{rot}$ from measured torsion pendulum
free-oscillation decay, characterized by energy decay time constant
$\tau_e$ or, equivalently, the quality factor $Q$,
\begin{equation}
\frac{1}{\tau_e} = \frac{2 \pi f_0}{Q} = \frac{\beta_{rot}}{I}
\: \: ,
\label{tau_e_beta}
\end{equation}
where $f_0$ is the pendulum resonant frequency and $I$ is its moment
of inertia.

\begin{figure}[tb]
\includegraphics[scale=0.6]{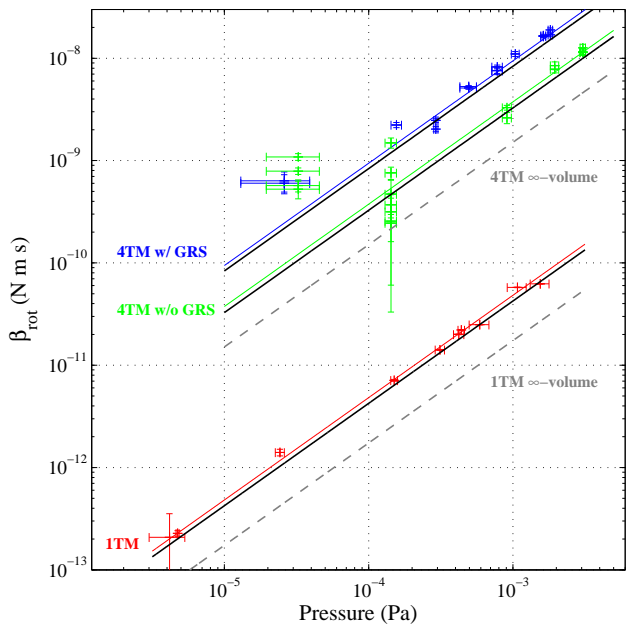}
\caption
{Plot of experimental data from the UTN gas damping experiment
\cite{utn_prl_damp} featuring cubic test masses in two different torsion
pendulums dominated by rotational (1TM) and translational (4TM)
squeezing, the latter with two different conditions of proximity to the
surrounding apparatus. The pressure independent 
damping has been removed from each dataset to show only the gas contribution.  
Linear fits to each dataset are shown. The
simulation predictions for the three cases are shown as thick black
lines, while the infinite volume limit predictions for the two pendulums
are shown in dashed gray.}
\label{utn_exper}
\end{figure}

Figure \ref{utn_exper} displays the data for the measured rotational gas
damping coefficient $\beta_{rot}$ from the University of Trento (UTN)
torsion pendulum ringdown measurements\cite{utn_prl_damp} with two
pendulum geometries featuring cubic TM (46~mm side).
Proximity-enhanced gas damping is studied in the geometry relevant to
the LISA gravitational wave observatory, where the TM is enclosed by a
rectangular capacitive sensor electrode housing -- referred to here as
gravitational reference sensor (GRS) -- with 3-4 mm
gaps\cite{utn_prl_damp}. In the first configuration (1TM), the cubic TM
is suspended on axis and thus a pure rotational damping is measured. In
the second configuration (4TM), four TM are each displaced by roughly
11~cm from the pendulum rotation axis in a cross configuration. This
results in both translational and rotational damping inside electrode
housings that surround two of the TM (one housing is the GRS mentioned
above and the other is a larger housing with 6-8~mm gaps). Measurements
with the 4TM were also performed after removing the GRS housing (labeled
``w/o GRS'' in the plot), in order to isolate its proximity effect on
the total damping. From each dataset the 
zero-pressure background pendulum damping has been subtracted off  
using a linear fit. This is
nearly irrelevant for the 1TM data, which employs a low damping fused
silica fiber\cite{utn_cqg_fused} -- $Q \approx 10^6$ -- while for the
4TM experiment, using tungsten fibers with $Q \approx $ 2000 - 3000, the
fiber damping is ten times larger than the gas damping contribution at
the lower pressures of the study. Here, uncertainty in the residual
damping is the dominant source of error and causes the large spread in
the data (several points with error bars overlapping zero have been
omitted from the log-scale plot, but have been used in obtaining the
fits to the 4TM data).

The UTN data demonstrate the proportionality of gas damping to pressure,
for more than three decades in pressure 
in the 1TM study and more than a decade in pressure in the
4TM study, with the wide dispersion mentioned above limiting the
validity at lower pressures. All the data demonstrate a clear -- factor
3 to 6 -- excess above the predictions of the infinite volume model
calculated from Ref.\cite{utn_inf_volume}. Additionally, the measurable
decrease in gas damping for the identical pendulum upon removal of the
GRS in the 4TM pendulum can only be attributed to proximity damping. The
three datasets are consistent, at the 20\% level, with simulations
that employ the same technique as that described for the LIGO
configuration in Sec. \ref{sec_numerical}, with the torque damping
coefficient calculated from the simulation torque fluctuations. Though
based on the same basic cubic geometry and gap sizes, the different
measurements probe both the translational and rotational aspects of
squeezing, which have slightly different dependence on gap. 

Not shown here, the UTN experiments \cite{utn_prl_damp} also verified
the Brownian nature of the force fluctuations associated with the gas
damping, detecting a frequency independent increase in the pendulum
torque noise floor consistent with $S_N = 4 k_B T \beta_{rot}$.

\begin{figure}[tb]
\includegraphics[scale=0.6]{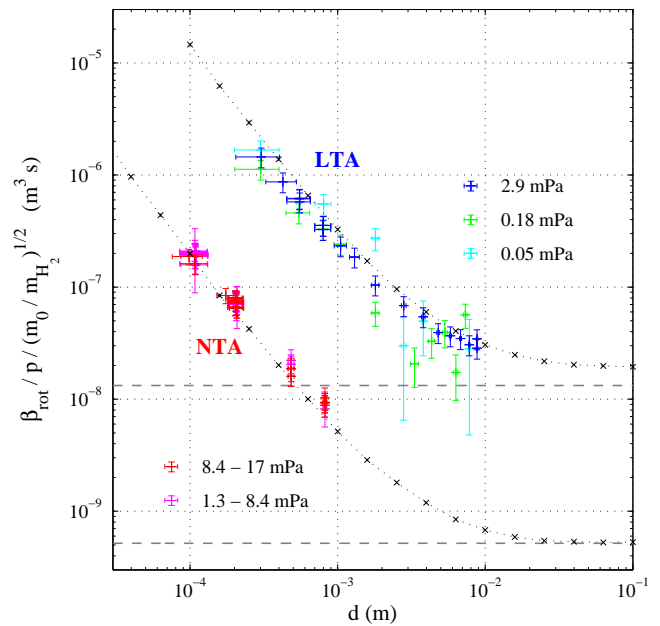}
\caption
{Plot of experimental data from the UW gas damping
experiment\cite{UW_squeeze_damping} for two different torsion pendulums,
labeled NTA and LTA, both with a rectangular plate geometry in proximity
of a second parallel plate at distance $d$. The pressure independent
damping has been removed from each dataset to show only the gas
contribution. Simulation points are shown with black $\times$, with
infinite volume limits shown in dashed gray.  Dotted lines connecting
simulation points are included as a guide for the eye.}
\label{uw_exper}
\end{figure}

The University of Washington (UW) experiments \cite{UW_squeeze_damping},
with results shown in \figr{uw_exper}, measure the gas pressure ringdown
damping of a two different torsion pendulums. Both inertial members,
labeled NTA and LTA, are thin rectangular plates in close proximity to a
second parallel plate, with the NTA dimensions 42~mm $\times$ 31~mm
$\times$ 2~mm and the LTA dimensions 114~mm $\times$ 38~mm $\times$
450~$\mu$m. Separations from 100~$\mu$m to 1~cm are studied, covering a
range of aspect ratios -- $b$/$d$, with $b$ the shorter dimension of the
rectangular plate -- of 4 to 400, thus well bracketing the LIGO
geometry, where $2R$/$d$~$\approx$~70. The data have been normalized for
linear pressure dependence, as well as for the different atomic masses
of the residual gas species, mostly N$_2$ and H$_2$O, using $\beta \sim
p \, m_0^{1/2}$. For each dataset the pressure independent damping has
been estimated, and removed, using the zero-pressure intercepts from
linear fits to groups of damping measurements performed at fixed gap
$d$. 

In addition to distributing statistically around a single
curve following normalization for pressure, the data demonstrate
the increase in damping with decreasing gap, by roughly two orders of
magnitude for a factor 30 decrease in gap for the LTA data. Also shown
with the data in \figr{uw_exper} are the predictions from the torque
noise simulation performed for this geometry as in Sec.
\ref{sec_numerical}. The simulation $\beta_{rot}$ agrees with the experimental
data to within 30\% across two decades of $d$, with residual 
differences considered compatible with systematic and statistical 
uncertainties in assigning pressures and gas species.   

Even for the largest gaps studied, $b$/$d$ $\approx$ 4 for LTA,
excess proximity damping is dominant over the infinite volume
contribution. The simulation for large $d$ is 
complicated by the presence of additional fixed control surfaces,
adding proximity damping that is independent of $d$\cite{UW_lisa_symp}. As
seen in \figr{uw_exper}, inclusion of geometrically simple control
electrode surfaces in the LTA simulation create a relevant increase of
the large $d$ predictions above the infinite volume 
prediction
\footnote{The expression for the torque damping coefficient
for a plate of width $a$, height $b$ and thickness $t$,
 can be found by simple integration of the 
force noise per unit area on the six faces of the prism, following Eqn. 14 
in \cite{utn_inf_volume}.  In the limit of vanishing prism thickness 
($t \ll a$), 
valid to better than 5\% for the geometries 
considered here\cite{UW_squeeze_damping}, 
$\beta_{rot} = \frac{p}{v_T} \left( \frac{1}{18 \pi} \right)^{1/2} b a^3  
\left( 1 + \frac{\pi}{4} \right)$. }.
Thus, while the experimental data are consistent with the simulation at total
damping values only several times larger than the infinite volume limit,
they do not allow a quantitative comparison with the infinite volume 
predictions. 

The UW results are incompatible with an elastic, specular scattering
description of the molecule - wall interaction. With elastic scattering,
the molecule velocity parallel to the plate surfaces is unchanged
as it bounces through the gap, and so the effective escape time $\tau$
becomes independent of gap. Following Eqn. \ref{fluc_diss_eqn_tau}, 
this would give force noise $S \sim \beta \sim d^{-1}$. The
steeper gap dependence observed in \figr{uw_exper}, with logarithmic
slopes of approximately $-1.5$ in the decade of smallest gap in both
configurations, is thus incompatible with an elastic scattering
scenario. Along with the success of the simulations in explaining the
data, at least for the systems under study here -- involving gold coated
surfaces and mostly water vapor and nitrogen residual gas at room
temperature -- the diffuse inelastic scattering hypothesis is well
verified. 
 
Taken as a whole, the two experimental campaigns measure proximity-enhanced
gas damping at levels that agree with simulation predictions to better than
30\% over several orders of magnitude in both pressure and gap size. 
Such a quantitative test of the physics and simulation techniques employed 
here for Advanced LIGO, in similar macroscopic systems, 
strengthens the confidence in the predictions for the gravitational 
wave sensitivity, which are discussed in the next section.

\section{Impact on gravitational wave detection capabilities and discussion}
\label{sec_discussion_results}

The data for $S_{F_0}$ and $\tau$ with $d$ = 5~mm, in Figs. \ref{S_F_d}
and \ref{tau_d} correspond to the current Advanced LIGO design geometry.
The low frequency noise power increase (or equivalently, the increase in the
gas damping coefficient $\beta$), with respect to the infinite volume 
limit is near 40, or slightly more than 6 in linear force noise 
spectral density. 

\begin{figure}[t]
\includegraphics[scale=0.52]{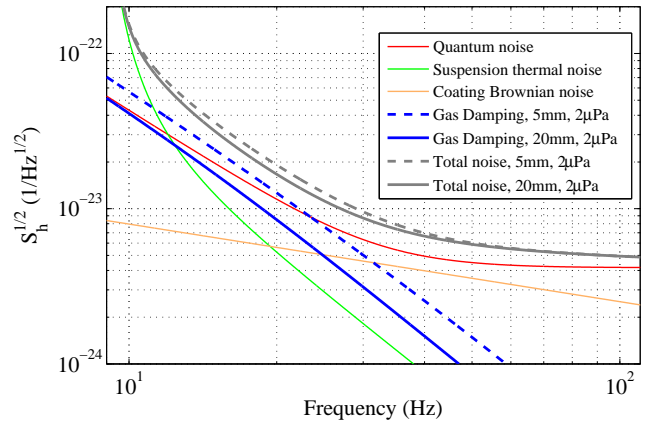}
\caption
{Gravitational wave strain measurement noise contributions for 
Advanced LIGO\cite{adv_ligo_spec}.
Dashed and solid blue curves correspond to the noise contributions from
gas damping noise with $2 \times 10^{-6}$ Pa of H$_2$ 
for the nominal geometry (all 5~mm gaps between all four 
TM - reaction masses pairs) and the alternate geometry with a 20 mm gap 
adjacent to the two input TM.  The total measurement noise in the two 
geometries are shown with thick gray curves.  
}
\label{fig_aLIGO_sens}
\end{figure}

Considering a design gas pressure of 2$\times 10^{-6}$ Pa (1.5 $\times
10^{-8}$ Torr) of H$_2$, this contributes a white low frequency force
noise $S_{F_0}^{1/2} \approx$ 20~fN \rthz, or an acceleration noise of
nearly 0.5 fm/s$^2$\rthz. This acceleration noise is assumed to act
equally on each of the four TM in the two interferometry arms. Converted
into an equivalent gravitational wave strain noise with 4~km
interferometry arms, the white force noise assumes an approximate
$f^{-2}$ shape in $S_h^{1/2}$ at low frequencies, and, as shown in
\figr{fig_aLIGO_sens}, rivals quantum fluctuations as the leading noise
source in the 10 - 30~Hz band \cite{adv_ligo_spec}\footnote{The
quantum noise estimate in Ref.~\cite{adv_ligo_spec} differs slightly
from that shown in \figr{fig_aLIGO_sens}, 
due to slightly different choice of tunable 
interferometer detection parameters.}.  This can have a
significant impact on the detection horizon for observing neutron star
inspirals, even more so for the merger of more massive black hole
sources, which have more signal contribution at the lowest frequencies.


We note that the calculated diffusion time is fast enough, 700 $\mu$s,
that the corresponding 3~dB rolloff frequency of roughly 230~Hz is above
the band where gas damping is critical to the detector sensitivity, with
photon shot noise limiting sensitivity at these frequencies for the
advanced detectors. 
 
The non-trivial impact of this noise source on the Advanced LIGO
sensitivity has raised the attention of the
collaboration\cite{norna_tech}, and we consider here a few
options for mitigation. For the central input TM (ITM), increasing the
gap between ITM and the reaction mass to 20~mm is feasible, which
reduces the associated force noise power on the ITM by roughly a factor
7 (see Figs. \ref{S_F_d} and \ref{spectrums}). The ITM reaction masses,
called compensation plates, are used for correcting the beam wavefront
for thermal lens effects and for relatively low force authority
electrostatic actuators. The reaction masses associated with the distant
end TM (ETM), on the other hand, are responsible for higher authority
electrostatic force control of the interferometer cavity length, and
increasing the gap would create a serious limit on the maximum possible
force. As such, the current baseline design maintains 5~mm gaps for the
ETM. Mitigation for the ETM force noise could be achieved by improving
the vacuum pressure at the end stations or, if necessary, adopting a
more complicated, possibly ring-like, geometry for the electrostatic
actuation reaction masses. The gas damping and total gravitational wave
strain with the increased ITM gaps is also shown in
\figr{fig_aLIGO_sens}.

It is worth reconsidering one initial assumption used in this analysis,
namely that the steady state pressure in the gap is equal to that of
the surrounding gas reservoir. In practice, any local
outgassing from the surfaces inside the gap will create a net radial
molecular flux and thus a local pressure increase above that of the 
surroundings. This increased effective pressure in the gap increases the
corresponding force noise, and thus maintaining the predictions obtained
for 2 $\mu$Pa of H$_2$ limits the tolerable outgassing from the test and
reaction mass surfaces. The pressure increase from a total gas
outgassing flux $Q$ (measured in [Pa\,m$^3$/s]) can be estimated from the
macroscopic flow estimates
of Sec. \ref{sub_sec_piston}, 
\begin{equation}
\Delta p = Q Z_M = Q \frac{\tau}{\pi R^2 d}
\: \: .
\label{eqn_pressure_rise}
\end{equation}    
Given the estimated values of $\tau$ in Sec. \ref{sec_numerical},
maintaining $\Delta p$ well below 2~$\mu$Pa in the 5~mm gap will require
outgassing well below 10$^{-6}$ Pa m$^3$/s. Per surface area, this
corresponds to 10$^{-8}$ mBar\,l/s/cm$^2$, using more familiar vacuum
technology units. This represents an important -- but, with reasonable
pumping times and temperatures, feasible -- requirement on outgassing
performance, which is likely to be dominated by water outgassing from
the optical surfaces\cite{schram,lewin_book}. 

In closing, we discuss a series of design concerns relevant to other
observatories with goals similar to those of Advanced LIGO, such as
Advanced Virgo\cite{adv_virgo_baseline_design} and the Large-scale
Cryogenic Gravitational wave Telescope (LCGT \cite{LCGT_cqg_2009, 
LCGT_cqg_2010}), and
to more ambitious future facilities, such as the Einstein Telescope (ET
\cite{ET_design}). The first issue is geometrical. Instead of simply
increasing the distance between similar shaped test and reaction masses
as in Advanced LIGO, which improves the linear noise spectral density
roughly proportionally to $d$, the proximity enhancement can be 
effectively removed 
for a reaction mass in the form of an external coaxial ring around the
cylindrical TM. In this geometry, TM motion along the cylinder (and
GW-measurement) axis does not compress the gas between the TM and
reaction ring. As such there is no gas flow in the gap and no related
excess proximity dissipation. The same physics
explains the absence of excess force noise perpendicular to the cylinder
axis in the Advanced LIGO geometry seen in Sec. \ref{sec_numerical}. 

Advanced Virgo employs such an external ringlike reaction mass. There
is an additional cylindrical thermal compensation 
plate facing the TM cylinder, but at a
distance of 20~cm \cite{adv_virgo_baseline_design} and thus virtually
irrelevant for its proximity (see \figr{S_F_d}). Considering the vacuum
requirements, dominated by 10$^{-7}$ Pa of H$_2$O, and TM dimensions
similar to those of Advanced LIGO, one expects a number several
times below that given in \eqr{S_F_inf}, and thus not a critical issue
for the target sensitivity. 

Proximity concerns aside, gas damping can be reduced by acting on the
factors relevant to the infinite volume limit, concerning the TM thermal
vacuum environment, which enter as a factor $p^{1/2} T^{1/4} m_0^{1/4}$
in linear spectral density, and the TM geometry, entering roughly as the
ratio of the square root of the the surface area ($R$) to mass ($R^2 h$).
LCGT will employ cryogenically cooled TM, at 20~K, with expected
pressures near 2$\times 10^{-7}$~Pa dominated by H$_2$. LCGT will thus
enjoy a rough factor 6 noise improvement over Advanced LIGO 
before considering
geometry (LCGT will also likely employ a ring geometry for the reaction
masses\cite{yamamoto_priv}). Though the TM are slightly smaller than
those of Advanced LIGO (30~kg of sapphire rather than 40~kg of fused
silica), there should be a net improvement with LCGT and thus gas
damping should not be critical for their design sensitivity. 

Several cryogenic issues complicate the gas damping analysis. The
molecules striking the TM are assumed to have been thermalized by 
a cryogenic thermal shield, but some alteration of the
velocity distribution by thermal gradients is still possible. Low
temperatures should enable lower pressure with cryopumping, freezing out
all relevant species except H$_2$. More delicate, but potentially
helpful, is the question of H$_2$ sticking on the TM. Sticking times
for H$_2$ on cold solid surfaces rapidly lengthen with decreasing
temperature, with a strong substrate dependence, from the $\mu$s - ms
range at 20 K to the 1~s - 1~year range at 10~K (see App.
\ref{app_sticking_time}). While we do not have published absorption
parameters for H$_2$ on the most probable oxide TM coatings, such as
SiO$_2$ and Ta$_2$O$_5$, it is likely that sticking times can rival or
exceed time-of-flight to surrounding surfaces (of order 20~$\mu$s for
5~mm at 20~K). This lengthens the relevant escape time $\tau$ and
progressively lowers the maximum frequency for which proximity effects
are relevant, even to below the band of interest\footnote{Even in the
infinite volume limit, there is a small decrease in noise at frequencies
above $\tau_{st}^{-1}$, as the correlation between the momentum exchange
from the arrival and reemission of a molecule striking the TM surface
disappears on such timescales. For the Advanced LIGO TM, the reduction
is roughly 20\%, see Table~1 and Section 2.1.1 of Ref.
\cite{utn_inf_volume}}. 

The Einstein Telescope, which has a dedicated low frequency
interferometer, aims at sensitivities of roughly 10$^{-24}$\rthz at
10~Hz with an increased armlength of 10~km and thus requires a factor 30
improvement over the Advanced LIGO acceleration noise. This will require
both an improved geometrical configuration, such as the just discussed
ring reaction mass design, and some combination of improved thermal
vacuum conditions and larger TM. The specified conditions (10$^{-8}$~Pa
and 10~K) alone give a factor 30 improvement over the same values for
Advanced LIGO. TM of roughly 200~kg and 500~mm diameter will give an
extra factor 3 improvement for the infinite volume prefactor. Assuming
that the target pressure is reached, gas damping does not appear to be
a dominant noise source for ET.

\appendix

\section{Estimation of molecular sticking time}
\label{app_sticking_time}

The sticking time, which we will call $\tau_{st}$, or average time that
a molecule spends on the surface of the test mass or adjacent apparatus
between impact and reemission, can be evaluated via its influence on the
two dimensional equation of state of the adsorbed film in equilibrium
with its gas vapor. This is characterized by number density $n_{2D}
\left( p, T \right)$. In equilibrium, the rate of molecules (per area)
that desorb from the surface to enter the gas phase,
$\frac{n_{2D}}{\tau_{st}}$, must be equal to the number of gas molecules
that strike the surface\cite{milton_book}. Integration over the
Maxwell-Boltzmann velocity distribution gives the latter, and we
find\footnote{Our analysis differs slightly here from that of Ref.
\cite{milton_book} in that we lump their probability of sticking ($s$,
pg. 110) into the sticking time to form a single sticking time which
represents the average time that a molecule spends, per collision, on
the surface.}

\begin{equation}
\frac{p}{\left( 2 \pi m_0 k_B T \right)^{1/2}} 
= \frac{n_{2D}}{\tau_{st}}
\:\: .
\label{eqn_equilib_gas_film_stick}
\end{equation}

For the low gas pressures and relatively high temperatures of interest
to our experimental system, we consider the adsorbed molecules as an
ideal non-interacting 2D gas in the $xy$ plane, with a substrate
potential $V \left( z \right)$ in the dimension perpendicular to the
surface. Approximating the surface potential with a well depth $D$ and
elastic constant $m_0 \omega^2$, this gives energy levels 
\[
\epsilon =
\frac{\hbar^2 k^2}{2 m_0} - D + \hbar \omega \left ( n +
\frac{1}{2}\right)
\]
 where $k$ represents the 2D wave vector. 
In the classical limit of low occupation number, this yields the following
equation of state 
\begin{equation}
n_{2D} = \frac{p}{k_B T} \left( \frac{2 \pi \hbar^2}{m_0 k_B T} \right)^{1/2}
\exp \frac{D}{k_B T}
\frac{1}{2 \sinh \frac{\hbar \omega}{2 k_B T}}
\:\: .
\label{eqn_state_2D}
\end{equation}

Combining this with \eqr{eqn_equilib_gas_film_stick}, we obtain 
\begin{equation}
\tau_{st} = \frac{2 \pi \hbar}{k_B T} \exp \frac{D}{k_B T} 
\frac{1}{2 \sinh \frac{\hbar \omega}{2 k_B T}}
\:\: .
\label{eqn_tau_stick}
\end{equation}
An intuitive, semi-classical interpretation of this sticking time 
emerges in the limit of classically excited oscillator states, with 
$\hbar \omega \ll k_B T$, which yields
\begin{equation}
\tau_{st} 
\approx 
\frac{2 \pi}{\omega} \exp \frac{D}{k_B T} 
\:\: .
\label{eqn_tau_stick}
\end{equation}
This limit corresponds to a molecule oscillating around the bottom of 
the potential well, ``attempting escape'' once per classical oscillation, 
with an Arrhenius escape probability of $\exp \frac{-D}{k_B T}$. 

Substituting a harmonic well for the true Van der Waals potential with
Coulomb repulsion is an approximation, both for the level spacing and
for the finite number of bound states. Additionally, this model treats
possible rotational excitations of the molecule as equally populated in
the gas and adsorbed phases. As such, we use this model only to have an
order of magnitude estimate of sticking times, to give a comparison with
the typical times of flight between collisions on adjacent test mass and
reaction mass surfaces. 
 
Typical measured values for the adsorption well depth of H$_2$ on
various solid substrates, including both dielectrics and metals, range
from 300 to 600~K, with oscillator level spacings of order several
hundred K \cite{vidali_adsorption_potentials}. This results in room
temperature sticking times of order 1~ps. The distinction between the
classical and quantum excitation levels of harmonic oscillator is a
minor correction to the sticking time estimates here, and the sticking
times correspond to several to tens of classical oscillation cycles.
More easily polarized molecules such as N$_2$, CH$_4$, and H$_2$O can
find adsorption well depths of order 1000~K with slightly lower
vibrational energies, resulting in sticking times of order 10-100~ps.
For all realistic residual gas species, the room temperature sticking
times are sub-ns, well below the typical times for a gas molecule to
cross a 5~mm gap, which is of order 5~$\mu$s for H$_2$ and longer
for larger molecules. As such, the sticking time of residual gas
molecules in Advanced LIGO is not expected to create a relevant
departure from the time dependence of force fluctuations given using the
approximation of immediate reemission of incoming gas molecules. 

For cryogenic temperatures, such as the 20~K foreseen for LCGT, the
situation is quite different. Bound molecules are frozen into the ground
state of the adsorption potential well, with a binding energy $\left( D
- \frac{\hbar \omega}{2} \right)$. The sticking time for H$_2$ on the
dielectric surface of foreseen oxide coatings such as SiO$_2$ or
Ta$_2$O$_5$ becomes very sensitive to the adsorption parameters -- for
which we do not have published values -- and the exact
temperature. For instance, for H$_2$ on the oxide MgO, with binding
energy 330~K, the estimated sticking time is of order 0.5~ms, with a
10\% change in either binding energy or temperature producing a factor 5
change in the sticking time. For H$_2$ on NaCl (280~K binding energy)
and on graphite (440~K), the sticking times are 3~$\mu$s and 7~ms. At
10~K, the range of $\tau_{st}$ for these species becomes 10~s to 1~year. 

In the Advanced LIGO-like geometry at cryogenic temperatures, it is thus
conceivable that the sticking time will be longer than the time for
molecules to cross the intercylinder gap. Though this complicates the
analysis, the general effect on gas damping can be understood with the
force shot noise model discussed in Sec. \ref{sub_sec_shot}. At a given
$p$ and $T$, sticking will not change the total momentum exchange
associated with a molecule's stay inside the gap or the rate at which
molecules enter the gap, but it will lengthen the typical time spent
inside the gap by $N \, \tau_{st}$ ($N$ is the typical number of
collisions that a molecule makes before diffusion out of the gap, Sec.
\ref{sec_analytical}, of order 100 for 5~mm gaps in Advanced LIGO). As
such, the low frequency force noise should be unchanged by sticking, but
the longer diffusion or correlation time results in a lower frequency
cutoff for the excess noise from any proximity effects. This could push
any excess proximity damping to frequencies below the band of interest.

\section{C\lowercase{alculation of} $\langle \Delta q^2 \rangle$}
\label{app_calc_dp2}
For a zero mean, statistically stationary 
random variable $F \left( t \right)$ with mean defined over 
some time interval $\left[ 0 , T_0 \right]$, 
$\bar{F} \equiv \frac{1}{T_0} \int ^{T_0}_0 F \left( t \right) dt$,
the expected fluctuations in $\bar{F}$ can be calculated in terms
of the autocorrelation function $C_{FF} \left( \delta \right) \equiv 
\langle F \left( t \right) F \left( t + \delta \right) \rangle$,
\begin{equation}
\langle \bar{F}^2 \rangle
= 
\frac{1}{T_0}
\int_{-T_0}^{T_0}  \: C_{FF} \left( \delta \right) 
\left( 1 - \frac{\left| \delta \right|}{T_0} \right)
\, d\delta
\:\: .
\label{app_F_ms}
\end{equation}
In our case, $F \left( t \right) $ is the instantaneous force
felt by the TM due to molecular impacts.  

The single-sided power spectrum $S_F \left( \omega \right)$ 
is defined by the Fourier transform of $C_{FF} \left( \delta \right) $,
$S_F  \left( \omega \right) = 2 \int^{-\infty}_{\infty} 
C_{FF} \left( \delta \right) \exp \left( -i \omega \delta \right) \, d\delta$.  
For a process characterized by white noise passed through an
 effective low-pass filter, such as our approximation of the gas
damping system studied here, with 
$S_F = \frac{S_0}{1 + \left( \omega \tau \right)^2}$, the 
corresponding autocorrelation function can be calculated to be
\begin{equation}
C_{FF} \left( \delta \right) 
= 
\frac{S_0}{4 \tau} \exp \left( -\frac{\left| \delta \right| }{\tau} \right) 
\:\: .
\label{app_autocorr_lowpass}
\end{equation}

Inserting this result into \eqr{app_F_ms} and integrating,
we obtain 
\begin{equation}
\langle \bar{F}^2 \rangle 
= 
\frac{S_0}{2 T_0} 
\left[ 
1 - \frac{1 - \exp \left( -\frac{T_0}{\tau} \right)}{ \frac{T_0}{\tau}}
\right]
\:\: .
\label{app_F_ms_spec}
\end{equation}
As the total momentum exchange with the TM in a measurement time $T_0$
is simply $\Delta q = \bar{F} T_0$, we recover the result in
\eqr{eqn_exp_delta_p_2}. 

\section{C\lowercase{alculation of} $\langle r^2 \rangle$ 
\lowercase {and} $\langle t \rangle$ }
\label{app_calc_r2_t}
\begin{figure}[t]
\includegraphics{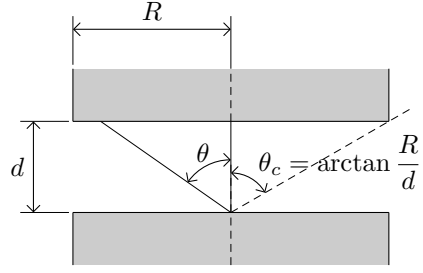}
\caption{Illustration of the geometry used for calculating the 
mean square lateral displacement and the mean time of flight for 
a molecule emitting into the gap from the center of one of the cylinders.
The angle $\theta_C$ represents the critical value of the polar angle
$\theta$, beyond which 
emission from the center results in the molecule escaping the gap laterally
in a single flight.  }  
\label{fig_x_RMS}
\end{figure}

Estimates for $\langle r^2 \rangle$ and $\langle t \rangle$ are obtained
by direct integration over the distribution of outgoing molecular
velocities for an emitted molecule 
\begin{equation}
P \left( \vec{v} \right) d^3 \vec{v} 
= 
\frac{\cos \theta}{2 \pi v_T^4}   
\, v \, \exp \left( - \frac{v^2}{2 v_T^2} \right)
\left( v^2 dv d\Omega \right)
\label{app_distrib}
\end{equation}
where $\theta$ and $\phi$ are the polar and
azimuthal emission angles. For both calculations, we divide the
distribution into a range of polar angles $\theta$ (see
\figr{fig_x_RMS}) for which a molecule emitted from the center of a TM
face strikes the opposing cylinder ($\theta < \theta_c$) and another for
which the molecule escapes the gap directly ($\theta > \theta_c$), with
the critical angle defined by the aspect ratio, $\tan \theta_c \equiv
\frac{R}{d}$. 

For the average time of flight, we obtain
\begin{eqnarray}
\langle t \rangle 
& = &  
\int_{\cos \theta_c}^{1} P \left( v,\theta \right) \frac{d}{v \cos \theta} 
v^2 dv d\Omega
\nonumber \\
\, & \, & +   
\int_{0}^{\cos \theta_c} P \left( v,\theta \right) \frac{R}{v \sin \theta} 
v^2 dv d\Omega
\nonumber \\
\, & = & 
\left( \frac{\pi}{2} \right) ^{1/2} \frac{d}{v_T}
\left[ 
 1 + \frac{R}{d} 
 \left( 1 - 
 \sqrt{ 1 + \left( \frac{d}{R} \right)^2}  \right) \right]
\nonumber \\
\, & \approx & 
\left( \frac{\pi}{2} \right) ^{1/2} \frac{d}{v_T}
\left( 1 - \frac{d}{2 R} \right)
\:\: .
\label{app_exp_t}
\end{eqnarray}

For the calculation of $\langle r^2 \rangle$ only the angular distribution
$P \left( \theta, \phi \right) d\Omega = \frac{1}{\pi} \cos \theta \,
d\Omega$ is relevant. We note that those molecules, emitted for 
$\theta > \theta_C$, that directly leave the gap on a single jump from the
center are weighted with a lateral displacement that is ``saturated'' to
be $R$. We obtain

\begin{eqnarray}
\langle r^2 \rangle 
& = &  
\int_{\cos \theta_c}^{1} P \left(\theta , \phi \right) d^2 \, \tan ^2 \theta \: 
d\Omega
\nonumber \\
\, & \, & +   
\int_{0}^{\cos \theta_c} P \left(\theta , \phi \right) R^2
 d\Omega
\nonumber \\
\, & = &
d^2 \ln \left[ 1 + \left( \frac{R}{d} \right)^2 \right] 
\:\: .
\label{app_MS_r}
\end{eqnarray}

We note that while the average time of flight remains finite for
cylinders of infinite radius -- $\langle t \rangle \to \left(
\frac{\pi}{2} \right)^{1/2} \frac{d}{v_T}$, as we use in the approximate
calculation (\eqr{RMS_x_t}) -- the mean square displacement diverges,
albeit slowly, as $ R/d \to \infty$. Saturating the maximum displacement
to $R$ makes physical sense and corresponds to the case of escape on a
single bounce. The logarithmic dependence of $\langle r^2 \rangle$ is a
fundamental difference with the result of Ref. \cite{suijlen_squeeze},
where the relevant random walk step size is taken to be $\langle r
\rangle \approx \frac{\pi d}{2}$ (and not the RMS $\sqrt{\langle r^2
\rangle}$ as used here). This would result in analogous expressions for $N$,
$\tau$, and $S_F$  similar to those expressed here (Eqns.
\ref{calc_N}-\ref{shot_noise_prediction}) but without the logarithmic
factor. The logarithmic factor, also relevant to similar expressions
for the molecular flow impedances of short pipes (for example Ref.
\cite{livesey}) is responsible for the nearly power law dependence of
$\beta$ (or $S_F$) on $1/d$, with an exponent that is not exactly 2, but
rather approaches 2, from below, only for vanishing gap, as observed
here and in the three other numerical and experimental studies cited
here (Refs. \cite{utn_prl_damp, UW_squeeze_damping, suijlen_squeeze}).

\begin{acknowledgments}
The authors would like to thank Norna Robertson and Jim Hough for
stimulating interest in the gas damping problem for Advanced LIGO. We
also thank David Shoemaker and Ray Frey for reviewing the manuscript.
LIGO was constructed by the California Institute of Technology and
Massachusetts Institute of Technology with funding from the National
Science Foundation and operates under cooperative agreement PHY-0757058.
The contribution from the Universit\`{a} di Trento was supported by the
INFN, ASI (LISA Pathfinder contract), and the Italian Ministry of
University and Research (PRIN 2008). The contribution from the group 
at the University of Washington was supported by NASA grant 
NNX08AY66G, by NSF grant PHY0969488, and by DOE funding 
for the CENPA laboratory. This article has been assigned LIGO 
document number p1100093.
\end{acknowledgments}


\end{document}